\documentstyle[12pt]{article}
\begin{document}
\def\strut{\rule[-.5cm]{0cm}{1cm}}
\def\dspace{\baselineskip = .30in}

\title{
\begin{flushright}
{\large\bf IFUP-TH 10/95}
\end{flushright}
\vspace{1.5cm}
\Large\bf INFLATION INDUCED SUSY BREAKING AND FLAT VACUUM DIRECTIONS}

\author{{\bf Gia  Dvali}\thanks{Permanent address: Institute of Physics,
Georgian Academy of Sciences,  \hspace{1cm}380077 Tbilisi, Georgia.
E-mail:
dvali@ibmth.difi.unipi.it }\\ Dipartimento di Fisica, Universita di Pisa
and INFN,\\ Sezione di Pisa I-56100 Pisa, Italy\\}

\date{ }
\maketitle

\begin{abstract} We discuss how the inflation induced supersymmetry breaking
affects the  flat directions of SUSY vacua. We show that under general
assumptions all gauge nonsinglet fields, parameterizing flat directions
(and in particular squarks and sleptons), get large $radiative$ masses
which are related to the value of the Hubble constant ($H$) and to the
expectation value of the inflaton field. This mass (typically $\sim H$) is
of ``non-gravitational'' origin and does not vanishes in the
global SUSY limit. Large radiative corrections are induced by $F$-term
(or $D$-term) density which dominates the inflationary universe and strongly
breaks supersymmetry. In such theories it is difficult to treat squarks and
sleptons as a light fields in the inflationary period.
 In the generic supergravity theories all flat directions, including
moduli, are getting curvature of order $H$. However, for
the  gauge-nonsinglet flat directions radiative contribution to the
curvature (induced by renormalizable gauge interactions) may be dominant.

\end{abstract}
\newpage

\dspace

\section{ Introduction}

 Characteristic feature of many supersymmetric theories is the existence
of the noncompact flat directions in the vacuum.
Such directions are parameterized by the vacuum expectation values
(VEVs) of the scalar fields, whose masses are induced only by the
supersymmetry breaking, which provides (typically small) curvature for the
flat vacua. Commonly accepted scenarios [1] assume that supersymmetry
breaking takes  place in certain isolated ``hidden sector'' and then
gets transfered to other sectors by some universal messenger interaction.
Resulting masses are of the order
\begin{equation}
m^2 \sim {|F|^2 \over M^2}
\end{equation}
where $F$ is a VEV of the $F$-term that breaks supersymmetry in the hidden
sector ($D$-type breaking is also possible) and $M$ is a scale of messenger
interaction. In the generic supergravity theories the messenger is assumed
to be gravity and thus $M = {M_{Planck} \over \sqrt{8\pi}}$ (reduced Planck
mass). Clearly, in this situation one has to arrange the theory in such a
way that $|F|^2 \sim M_WM$ (in the vacuum with zero cosmological constant),
in order to generate soft masses
$\sim M_W$ (weak scale) in the matter sector
(compatible with the solution of the hierarchy problem).

There is a large class of the flat directions, which are
parameterized by the fields having
only $nonrenormalizable$ ($M$ suppressed) interactions. Such fields are
usually called moduli. Due to the universal nature of the gravity transfer
supersymmetry breaking (1), the resulting masses of moduli are $\sim M_W$,
very much like the squarks and sleptons. This fact may lead to the grave
cosmological difficulties [2] due to the very late decay of moduli Bose
condensate. The problem (partially) results from the assumption that moduli
masses during inflation are $\sim M_W$ and, therefore, it becomes very
difficult to dilute such a condensate, unless the Hubble constant (H) is
$\sim M_W$. However, as we have shown recently [3], in generic supergravity
theories (independently from the details of the inflation) the moduli
masses are of order $H$ and as a matter of fact in wide class of theories
are larger than $H$. This has to do with the fact that $any$ inflation
breaks supersymmetry since it provides a large cosmological constant and
thus large $F$-term (or $D$-term) density. Such inflation induced SUSY
breaking stabilizes flat directions giving mass $\sim H$ to the
moduli. This fact can have an important consequences
for the cosmological moduli problem, since
now the moduli condensate can be eliminated by any inflation, provided
there is no large displacement of the moduli minimum (from its present value)
[3]. As we have shown, in particular this can be the case if
moduli has no couplings (other than canonical term) in the Kahler potential.

Besides, in the SUSY theories there are many other flat directions
whose zero modes carry ordinary gauge quantum numbers and/or have
renormalizable interactions with the other fields. Simplest example is
provided by some components of squarks and sleptons in minimal SUSY
standard model or GUTs[4]. Such flat directions are not problematic
cosmologically, but can be of certain importance for the baryogenesis[5].
Very much like moduli fields, gauge nonsinglet flat directions are getting
curvature of order $H$ by generic supergravity inflation. However, unlike
the moduli their masses get contribution from the other (nongravitational)
sources as well.
In the present paper we show that such contributions
in general are not less important and in many cases can be
even dominant. The generic reason is that for the gauge nonsinglet flat
directions there are several candidate forces which can transfer the
message about SUSY breaking. In particular such are the gauge interactions
and precisely this was the basic idea of the ``old'' globally supersymmetric
models with ``geometric hierarchy''[6]. In this models the message about
the SUSY breaking from the hidden sector (say GUT Higgses) was transfered
to the standard model fields radiatively,  by the gauge interaction.
Clearly, this is not
a case in the conventional supergravity theories in which the hidden
sector is assumed to be trivial under visible gauge symmetries [1].
The crucial point however is that, even if such mechanism is not operative
in the present vacuum (with zero cosmological constant), in general it
had to be effective in the early universe, since the hidden sector
$F$-term, which provides SUSY breaking ``today'', in general is not the
one that was dominating energy density in the inflationary universe.
As it is known [7], during inflation universe has to be in the state
with large cosmological constant, which may or may not be a local minimum
of the theory. However, the essential requirement is that classical
expectation values of the fields change slowly. Below we
will assume that this change is slow enough, so that it makes sense
to perform the quantum expansion in perturbation theory about the points
of inflationary trajectory (of course, if the inflationary state is a local
minimum, this assumption is automatically valid).

Under above assumption,
we show that strong corrections had to be presented in the large class of
scenarios, in which the inflaton sector is not isolated (at least) from
some gauge nonsinglet fields. They give universal (up to a gauge quantum
numbers) contribution $\sim H$ to the curvature of all gauge-nonsinglet flat
directions, which in some cases can dominate over the similar contribution
induced by supergravity transmitted SUSY breaking.
This corrections are induced
by the large SUSY breaking in the early universe and vanish (or become
negligible) once the system settles in to the present minimum with
``weakly'' broken supersymmetry.

\dspace

\section{ Globally supersymmetric inflation}

 In this section we will consider the behavior of the SUSY vacua flat
directions
during inflation in the globally supersymmetric theory. The effect is of
the special importance due to the fact that it persists also in the locally
supersymmetric generalizations, but since its origin is
``non-gravitational'', it is more convenient to study it first in
the global SUSY
case. So let us consider the system of $n$ chiral superfields $S_i$ (where
$i=1,2,....n$). The scalar component of each superfield
we will denote by the same symbol, whereas the fermionic and auxiliary
($F$) components we will denote as $\Psi_i$ and $F_i$ respectively.
We will assume that our theory is invariant under a simple gauge
group $G$ (in reality such factors can be several) and that some
superfield form its irreducible representations.
Now, in the case of global supersymmetry the scalar potential is given by
[8];

\begin{equation}
V = |F_i|^2 + (D-terms)
\end{equation}

Let us assume that, in some way, the above system appears in the inflationary
state. The basic idea of any inflationary scenario is that at some time
the universe is dominated by the large vacuum energy density $V$, from
which it follows that $any$ inflation in the SUSY framework implies large
$F-term$ or $D-term$ density. Below we will assume that $D-terms$
vanish and thus $F$-terms are dominating inflation. Let's take such
to be an $F_s$-term of $S$ superfield. Thus, in our system the supersymmetry
is broken in the inflationary period by the amount that is measured
by the expectation value of $Fs$. Clearly, in the state with such a strongly
broken supersymmetry, the flat directions of the SUSY vacua can easily
be shut down (or destabilized) and corresponding zero modes in general
can get large masses, provided the SUSY breaking is transmitted to the
respective sector by some interaction. However, in the case of global
SUSY there is no universal messenger interaction that can transmit
supersymmetry breaking from one sector to another. Therefore, the effects
under consideration can only take place if some general conditions
are satisfied. In particular, such condition is:

(*) The superfield $S$, whose $F_S$-term is dominating inflation, is
coupled to some of the gauge nonsinglet fields in the superpotential.

Let $\phi$ be an gauge nonsinglet field in the real (say adjoint)
representation of $G$ (alternatively one can consider two fields
$\phi, \bar {\phi}$ in conjugate
representations), which is coupled to $S$ in the superpotential
\begin{equation}
\Delta W = gS\phi^2
\end{equation}
(obviously $G$-invariant contraction of the indexes is assumed).
In such a theory, the all $G$-nonsinglet flat directions will obtain a
$radiative$ two loop masses of the order
\begin{equation}
m^2_{rad} \sim ({\alpha \over 4\pi})^2 g^2 {|F_S|^2 \over M_{\phi}^2}
\end{equation}
where $M_{\phi}$ is a supersymmetric mass term of $\phi$ and $\alpha$
is a gauge coupling of $G$. If (3) is the
only contribution to the mass, then
\begin{equation}
M_{\phi} = g S
\end{equation}
where $S$ is the expectation value of the scalar component during inflation.
Notice, that $S$ need not necessarily be the inflaton field.
Experts will easily recognize in (4) the well known expression for the
radiative two loop scalar mass in the ``old'' globally supersymmetric models
with ``geometric hierarchy''[6]. In this models such corrections where arranged
to take place in the phenomenological minimum (present vacuum) and they
were the major source for transfer of SUSY breaking to the visible sector
trough the gauge interactions. Naturally, in these schemes the $F$-terms were
restricted to be $\sim M_WM_{\phi}$, resulting in the radiatively induced
mass $m_{rad} \sim M_W$.

Our observation here is that, independently whether such corrections are
zero in the present vacuum, they had to be presented
during inflation (provided (*) is valid).
Once again, this is consequence of the large $F_S$ density
in the early universe. Of course, in reality the global minimum is ``slightly''
nonsupersymmetric, but this does not affects significantly present discussion.
Above correction induces large mass to all scalar fields parameterizing
$G$-nonsinglet flat directions (squarks sleptons) and in general to all
$G$-nonsinglet light fields as well. However, for the fields which do
not correspond to flat vacua, there can be a larger tree level contribution
from the other sources.

 Using relation between vacuum energy and the Hubble constant (in the slow
roll approximation) [7]
\begin{equation}
H^2 = {V \over 3M^2} = {|F_S|^2 \over M^2}
\end{equation}
we can rewrite (4) in the following form
\begin{equation}
m_{rad}^2 \sim H^2 ({\alpha \over 4\pi})^2 ({M \over |S|})^2
\end{equation}
This form shows that, in general, $m_{rad}$ can be even larger than the
Hubble constant, if $|S|$ is somewhat bellow $M$. Usually, $|S|$ changes
(slowly) during inflation and so does the ratio $m_{rad}/H$.

\subsection*{Example} Now we wish to demonstrate above effect on the particular
example of `hybrid' inflation which originally was considered by
Linde [9] in nonsupersymmetric framework and later was studied in the
supersymmetric context in [10,11]. This scenario automatically satisfies
condition (*), since it implies that inflaton couples to gauge nonsinglet
Higgs field. The simplest superpotential which leads to the hybrid
inflation is

\begin{equation}
W = {1 \over 2} gS\phi^2 - S\mu^2
\end{equation}

where $\mu$ is a some large mass scale and if $G$ is grand unification
symmetry, then one has to assume $\mu \sqrt{2 \over g} = M_{GUT}$.
In order to study inflationary dynamics, let us assume for a moment
that $\phi$ also is
a gauge singlet field. Then, the scalar potential is given by

\begin{equation}
V = |F_S|^2 + |F_{\phi}|^2 = |{1 \over 2}g\phi^2 - \mu^2|^2 + g^2|S|^2|\phi|^2
\end{equation}

This theory has a unique supersymmetric vacuum with $\phi^2 = {2 \over g}\mu^2$
and $S=0$ in which all $F$-terms vanish. However, if we minimize $V$ with
respect to $\phi$ for the fixed values of $S$, we can easily
find that for $S > S_C = {\mu \over \sqrt{g}}$, the minimum is at $\phi = 0$,
potential is flat in $S$ direction and has $S$-dependent curvature
($\sim g|S|$) in
$\phi$ direction. Following [11], let us consider the chaotic
initial conditions
with $S >> S_C$. Since the curvature in the $\phi$ direction is very large,
we expect that $\phi$ will rapidly settle in its ``minimum'' with
$\phi = 0$. Contrastly, the curvature in $S$ direction is zero and, therefore,
the system can stay at $S >> S_C, \phi = 0$ quit long.
This state is dominated by large
$|F_s| = \mu^2 $ term density and inflation results. Notice, that
classically there is
no force that can drive $S$ to the global minimum. Such a force can
be provided by the positive SUSY-violating soft mass term $m^2|S|^2$ [10].
However, as it was shown in [11], $independently$ from the existence of the
soft terms, nonzero curvature of the $S$ slope in any case is provided
by one loop corrections to the effective potential[12]
\begin{equation}
\Delta V = {(-1)^F \over 64\pi^2} TrM^4ln{M^2 \over \Lambda}
\end{equation}
where the summation is over all helicity states, factor $(-1)$ stands for
the fermions and $\Lambda$ is a renormalization mass.
Crucial point is that in the region $S > S_C$ this corrections are nonzero,
even if the global vacuum of the theory is supersymmetric.
One may think that the existence of the nonzero
radiative corrections to the effective potential, in the theory with
supersymmetric ground state, may be somewhat surprising and contradict to
the ``nonrenormalization'' theorem [13]. Notice however, that
expansion is performed about the (classically flat) points $S > S_C$ and
$\phi = 0$ for which
supersymmetry is $broken$ by the large $F_S = \mu^2$ density and the Fermi Bose
masses are not degenerated. In the other words, in the early universe
the system is far away from its own global minimum and does not ``knows''
whether this minimum is supersymmetric or not. So, in this epoch there
are nonzero corrections. In fact, the nonexistence of this
corrections would be more surprising. Let us neglect gravity and inflation
for a moment and imagine that our system is at some point $S >> S_C$
and $\phi = 0$. Since
classically there is no driving force, the system can stay there long
enough, so that the hypothetic observer can measure the particle spectrum.
He would find: (1) one massless scalar $S$ and one massless fermionic
partner $\psi_S$ (goldstino); (2) one fermion $\psi_{\phi}$ with mass
$M_{\phi} = gS$ and two real scalars $\phi + \phi^*$ and $i(\phi - \phi^*)$
with $[mass]^2$s  $M^2_- =g^2|S|^2 - g \mu^2$ and
$M^2_+ = g^2|S|^2 + g \mu^2$ respectively.
Of course, with such a spectrum our observer can never conclude that
his universe is supersymmetric.

 One loop corrected effective potential (for $S >> S_C$) is given by [11]

\begin{equation}
V = \mu^4(1 + {g^2 \over 32\pi^2} [2ln{gx\mu^2 \over \Lambda^2} +
(x - 1)^2ln(1 - x^{-1}) + (x + 1)^2ln(1 + x^{-1})])
\end{equation}

where $x = {g|S|^2 \over \mu^2}$. In order to feel more comfortable with this
expression, we can obtain the same result by taking the exact SUSY
limit of the softly broken theory, for which
the point $S >> S_C, \phi = 0$ is a local minimum. For this, let us add
to the potential (9) the soft SUSY breaking terms.
\begin{equation}
V_{soft} = \epsilon^2 |S - |S_0||^2
\end{equation}
where $|S_0| >> S_C$. Now the theory has a well defined local minimum
at $S = S_0$ and $\phi = 0$ where curvature in all directions is positive.
The tree level spectrum in this minimum
is precisely the same as in the case $\epsilon = 0$ with only modification
that now inflaton $S$ gets a soft positive $[mass]^2 = \epsilon^2$.
The one loop
effective potential now is given by (11) plus an additional term proportional
to
\begin{equation}
V_{\epsilon} = \epsilon^4 ln {\epsilon^2 \over \Lambda^2}
\end{equation}
which vanishes in the limit $\epsilon \rightarrow 0$ and we are left with
(11). Above discussion is not altered significantly for the
gauge nonsinglet $\phi$,
since for $S >> S_C$ the gauge symmetry is unbroken. The phase transition
with gauge symmetry breaking takes place only after the $S$ field drops to its
critical value $S_C$. Below this point all the fields rapidly adjust their
VEVs in supersymmetric minimum. Inflation ends when the slow roll condition
breaks down and this happens when $S$ approaches $\sim S_C$ (from above).
We will not provide here all the details of this scenario for which
reader is referred to [11].  For us the most important thing about
this scenario is that
the $F_S$-term, which dominates the inflationary universe, is coupled
to the $G$-nonsinglet superfields and splits the masses of its
Fermi-Bose components. This results in to the large radiative masses of
all $G$-nonsinglet flat directions (and in particular quarks and
leptons) given by (7). In above model inflation ends when $S \sim S_C$
and thus the squark masses can be in general larger then the Hubble constant
(provided $\mu$ is small enough).

 The considered universal two loop radiative corrections to $[mass]^2$ are
positive and the flat directions are stabilized at the origin. However,
in some cases the dominating negative one loop corrections can
also appear. This may happen if the nonzero $F_S$-term is coupled to the
pair of fields $\phi, \bar {\phi}$ in the non-selfconjugate representations,
whose supersymmetric masses are splitted due to the mixing with other
(non-selfconjugate) representations. In general, such a situation may lead
to the nonzero one loop contribution to the $[mass]^2$, which are proportional
to the Abelian generators of $G$ (e.g. hypercharge in the GUT case) and
therefore, can have either sign. These corrections can destabilize flat
directions. Their presence in the present vacuum would be a disaster,
but in the early universe they can play very important role for the
baryogenesis via Affleck-Dine mechanism [5], since this mechanism requires
large expectation values of squarks and sleptons along the flat vacua.

 \section{Supergravity}

The effects considered in the previous section will be
operative in the supergravity scenarios as well if the large $F$-term
couples to some of the gauge nonsinglet fields.
However, as we have shown in [3], in the supergravity
framework flat directions (and in particular moduli) get curvature
$\sim H$ through the gravity transfered supersymmetry breaking.
Since the gravity is universal messenger, flat directions of the
both type (gauge-nonsinglets and moduli)
are in general disturbed by the equal strength
and resulting curvature is of the order
\begin{equation}
m^2 \sim {|F_S|^2 \over M^2}
\end{equation}
But, as we know, G-nonsinglet zero modes
can get extra contribution (4),(7) from the gauge interactions
which can be equally important.
Let us consider generic supergravity scalar potential [14]:
\begin{equation}
 V = exp({K \over M^2})[K^{i-1}_jF_iF^{*j} - 3{|W|^2 \over M^2}]
\end{equation}
where $K(S_i,S_i^*)$ is a Kahler potential, $W(S_i)$ is superpotential
and $S_i$ are chiral superfields. $F_i$ -terms are given by
$F_i = W_i + {W \over M^2}K_i$ where upper (lower) index denotes derivative
with respect to $S_i$ ($S_i^*$) respectively. ( Again, we neglect possible
$D$-terms and assume that they vanish during inflation).

 For simplicity we assume that flat direction mode (which we denote by $Z$)
enters in the Kahler potential only through the canonical term:

\begin{equation}
K = |Z|^2 + K'(S_i,S_i^*)~~~~~~ W = W_s + W'(S_i,Z)
\end{equation}

Where $K'$ and $W_s$ are arbitrary functions independent of $Z$.
Note that if $Z$ is moduli, then only Planck scale suppressed couplings
are allowed in $W'$.
G-nonsinglet zero modes are allowed to have renormalizable couplings in
$W'$.

 By convention, let us put the present minimum of the flat direction at
$Z = 0$ (at least for the squarks and sleptons this is the
necessary requirement).
In the class of models in which the minimum is not displaced during
inflation, the mass of $Z$-mode is given by:

\begin{equation}
 m_z^2 = 3H^2 +e^{K' \over M^2} [{|W|^2 \over M^2} + |W'_{zz}|^2 +
K'^{i-1}_j F_{iz}F^{*jz} - {|W'_z|^2 \over M^2}]
\end{equation}

where lower (upper) index $z$ denotes derivative with respect to $z(z^*)$.
The value of the second term in brackets is in general model dependent.
For the moduli (having only nonrenormalizable couplings in $W$) it can
be shown that, under some general conditions, this quantity is positive[3]
and thus in such cases moduli masses are larger than the Hubble constant.
In this situation inflation can dilute moduli condensate and solve the
problem.
For squark flat directions this contribution will again be positive if,
for example,
there is some symmetry (e.g. matter parity) for which $W'_z$ automatically
vanishes in the minimum with zero matter VEVs. The generic
message, however, is
that supergravitationaly induced curvature of the flat directions
is $\sim H$. One may compare this contribution with the one that can be
transfered by gauge interaction (7). We see that in general this two
sources are
comparable and gauge contribution may be even dominant if inflation
ends (slow roll condition breaks down) when $S$ is somewhat below
$M$.

\section{conclusions}
 In conclusion, we have shown that under general conditions all gauge
nonsinglet flat directions get large ($\sim H$) radiative contribution
to the curvature in the inflationary epoch. These corrections result from
the strong breaking of the supersymmetry, induced by the inflation, and
disappear after system adjusts to its present vacuum.
In generic supergravity theories all flat directions (including moduli)
are getting curvature $\sim H$ by the universal gravity transfered SUSY
breaking. For gauge non-singlet flat directions this two contributions
can be comparable and in some cases the gauge contribution may be
the dominant one.

\subsection*{Acknowledgments}
I would like to thank Riccardo Barbieri for very useful discussions.

  \section*{References}

\begin{enumerate}

\item R.Barbieri, S.Ferrara and C.A.Savoy, {\it Phys.lett.}, {\bf B119}
(1982) 343; A.H.Chamseddine, R.Arnowitt and P.Nath, {\it Phys.Rev.Lett.},
{\bf 49} (1982) 970;
L.Hall, J.Lykken and S.Weinberg, {\it Phys.Rev.} {\bf D27} (1983) 2359.

\item G.Coughlan, W.Fischler, E.Kolb, S.Raby and G.Ross, {\it Phys.Lett.},
{\bf B131} (1983) 59; J.Ellis, D.V.Nanopoulos and M.Quiros, {\it Phys.Lett.},
{\bf B174} (1986) 176; G.German and G.G.Ross, {\it Phys.Lett.}, {\bf B172}
(1986) 305; O.Bertolami, {\it Phys.Lett} {\bf B209} (1988) 277;
R. de Carlos, J.A.Casas, F.Quevedo and E.Roulet, {\it Phys.Lett.} {\bf B318}
(1993) 447;
T.Banks, D.Kaplan and A.Nelson {\it Phys.Rev}, {\bf D49} (1994) 779;
 T.Banks, M.Berkooz and P.J.Steinhardt, preprint RU-94-92.

\item G.Dvali, preprint IFUP-TH 09/95, hep-ph/9503259;

more recently another preprint has appeared, in which the same observation
was made, M.Dine, L.Randall and S.Thomas, preprint hep-ph/9503303.

\item e.g. see: E.Witten, {\it Nucl.Phys} {\bf B202} (1982)
253.

\item I.Affleck and M.Dine, {\it Nucl.Phys.} {\bf B249} (1985) 361.

\item J.Ellis, L.Ibanez and G.G.Ross, {\it Phys.Lett.} {\bf B113}
(1982) 283;
D.Polchinski and L.Susskind, {\it Phys.Rev.} {\bf D26} (1982) 3661;
L.Alvarez-Gaume, M.Claudson and M.B.Wise, {\it Nucl.Phys.} {\bf B207}
(1982) 96;
S.Dimopoulos and S.Raby, {\it Nucl.Phys} {\bf B219} (1983) 479;
M.Dine and W.Fischler, {\it Nucl.Phys.} {\bf B204} (1982) 346;

\item A.H.Guth, {\it Phys.Rev} {\bf D23} (1981) 347; For the review see
A.D.Linde, {\it Particle Physics and Inflationary Cosmology} (Harwood
Academic, Switzerland, 1990); E.W.Kolb and M.S.Turner, {\it The Early
Universe} (Addison-Wesley, Reading, MA, 1990);

\item For a review see e.g. H.-P.Nilles, {\it Phys.Rep} {\bf 110} (1984) 1.

\item A.D.Linde, {\it Phys.Lett} {\bf B259} (1991) 38;
{\it Phys.Rev.} {\bf D49} (1994) 748.

\item E.Copeland, A.R.Liddle, D.H.Lyth, E.D.Stewart and D.Wands,
{\it Phys.Rev.} {\bf D49} (1994) 417.

\item G.Dvali, Q.Shafi and R.Schaefer, {\it Phys.Rev.Lett.} {\bf 73}
(1994) 1886.

\item S.Coleman and E.Weinberg, {\it Phys.Rev.} {\bf D7} (1973) 1888.

\item M.T.Grisaru, W.Seigel and M.Rocek, {\it Nucl.Phys.} {\bf B159}
(1979) 429.

\item E.Gremmer, B.Julia, J.Scherk, S.Ferrara, L.Girardello and
P. van Nieuwenhuizen, {\it Nucl.Phys.} {\bf B147} (1979) 105-131.

\end{enumerate}
\end{document}